# Ultra-Low Noise Multiwalled Carbon Nanotube Transistors

Olli Herranen,[1][*] Deep Talukdar,[1,2][*] and Markus Ahlskog[1]

[1)]*Nanoscience Center, Department of Physics, University of Jyväskylä, P.O. Box 35, FI40014 Jyväskylä, Finland*

[2)]*Saha Institute of Nuclear Physics, 1/AF, Bidhannagar, Kolkata-700064, India*

We report an experimental noise study of intermediate sized quasi ballistic semiconducting multiwalled carbon nanotube (IS-MWCNT) devices. The noise is two orders of magnitude lower than in singlewalled nanotubes (SWCNTs) and has no length dependence within the studied range. In these channel limited devices with small or negligible Schottky barriers the noise is shown to originate from the intrinsic potential fluctuations of charge traps in the gate dielectric. The gate dependence of normalized noise can be explained better using ballistic the charge noise model rather than diffusive McWhorter's model. The results indicate that the noise properties of IS-MWCNTs are closer to SWCNTs than thicker MWCNTs. These results can be utilized in future to analyze noise in other purely ballistic nanoscale devices.

[*] corresponding author. E-mail: olli.herranen@jyu.fi

[*] corresponding author. E-mail: dtalukdar@ntu.edu.sg

# 1. INTRODUCTION

Singlewalled carbon nanotube (SWCNT) transistors working in the ballistic regime have generated considerable interest as building blocks of various nanoelectronic applications and sensors. However the performance limits for such nanoscale devices are set by their inherent low frequency noise. Though the low frequency noise characteristics in semiconducting ballistic SWCNT-FETs have been studied in detail by several groups [1-4], the studies have mostly been plagued by considerable Schottky barriers at the contact. No low frequency noise study was however forthcoming in channel-limited ballistic devices for SWCNTs. Such noise studies in channel limited devices can nevertheless be made in large diameter multiwalled CNTs owing to negligible Schottky barriers at the contacts. However, band gap in CNTs being inversely proportional to tube's diameter [5] ($E_g = \alpha/d$, $\alpha \approx 0.7$ eV nm), large diameter (> 10 nm) MWCNTs are predominantly metallic. That expression, though, is a crude estimate which gives the right order of magnitude but not the exact value for the band gap. Another hindrance for using large diameter tubes is disorder which leads to diffusive transport through the channel. Such situation can be dealt by using smaller diameter MWCNTs (diameters up to 10 nm) for performing noise studies in channel limited ballistic devices as they possess structure and properties reasonably close to SWCNTs [6, 7]. In the rest of the paper such tubes would be addressed as intermediate sized MWCNTs (IS-MWCNT).

Using IS-MWCNTs with short channel lengths compared to the inelastic scattering length and Palladium (Pd) as the contact metal we can obtain devices showing ballistic or quasi-ballistic transport having small or negligible Schottky barriers at the contacts. Pd contacts have been proved to form good contacts with the semiconducting nanotube (see e.g. ref. [8]) because of the good wetting properties and proper work function. For metallic nanotubes other materials such as titanium (Ti) and chromium (Cr) [9, 10] as well as Pd [11] can be used to obtain ballistic behavior. For such ballistic devices it is difficult to explain gate behavior of low frequency noise using conventional models and newer approaches such as charge noise model [12] or McWhorter's model [4, 13] is required to analyze their noise behavior. While most of these studies have been done in Schottky barrier [1] and liquid gated [14] SWCNT devices or *2-D* graphene transistors [15], noise behavior in mainly channel limited ballistic nanoscale transistors have not been studied. In case of quasi-ballistic MWCNTs, some noise studies exist [16] however its intrinsic origin noise and gate dependence were not studied. For large diameter diffusive MWCNTs several noise studies exist in literature [17-19].

In this work we investigate low frequency noise in semiconducting IS-MWCNT devices. We obtain normalized noise values for the devices to be lower than SWCNTs & MWCNTs and the noise dependence on gate consistent with the charge noise model [12]. In total four different MWCNT devices with tube diameters varying between $3-7$ nm and the channel lengths between $130-570$ nm were studied. The tubes were highly conductive reaching the ballistic limit [20-22] having the ON state resistances in the range of $R = 6.2 - 18$ kΩ (see table I). In general, the current through MWCNT is carried only by the two outermost shells [20, 23] and inner shells do not have any contribution for conductance and thus noise. A rough estimate for the upper limits of the band gaps for our measured tubes can be done by measuring the $I-V$ curve at a gate voltage corresponding to conductance minimum and determining the size of non-conducting region from it [24]. With such kind of an estimate we obtain gap values of few tens of meV i.e., well below 100 meV, closer to thermal energy at room temperature (25 meV) making the experiments possible only at low temperatures. The devices don't have much practical applications as FETs, however they provide important information on noise properties of channel-limited ballistic FETs.

## 2. EXPERIMENTAL

*Fabrication of IS-MWCNT devices:* The MWCNT devices were fabricated on top of highly doped Silicon wafer covered with 300 nm thick thermally grown $SiO_2$. Nanotubes (MWCNT material from collaborating group [25] & Sigma-Aldrich) were mixed into (1,2)-dichloroethane and the suspension was spin deposited on the substrate leaving tubes laying randomly on top of the sample surface. The tubes were located using AFM (atomic force microscope) and standard e-beam lithography steps were used to electrically contact the nanotubes. The contact metal was chosen to be Pd in order to improve the contact resistance. Schematic of the measurement setup is presented in figure 1.

*Measurement setup:* The transport measurements were performed in a RF-shielded room using homemade dipstick at 4.2 K. The noise measurements were performed by using a current preamplifier (Ithaco 1211) followed by subsequent digitization using a NI-DAQ device. The time series was further analyzed using MATLAB. Furthermore, analog low pass filters were used in the measurement chain to remove any aliasing effects. To obtain clean power spectra, background amplifier noise was subtracted. Further details of the measurement process can be obtained in ref. [26] and ref. [27].

## 3. RESULTS AND DISCUSSION

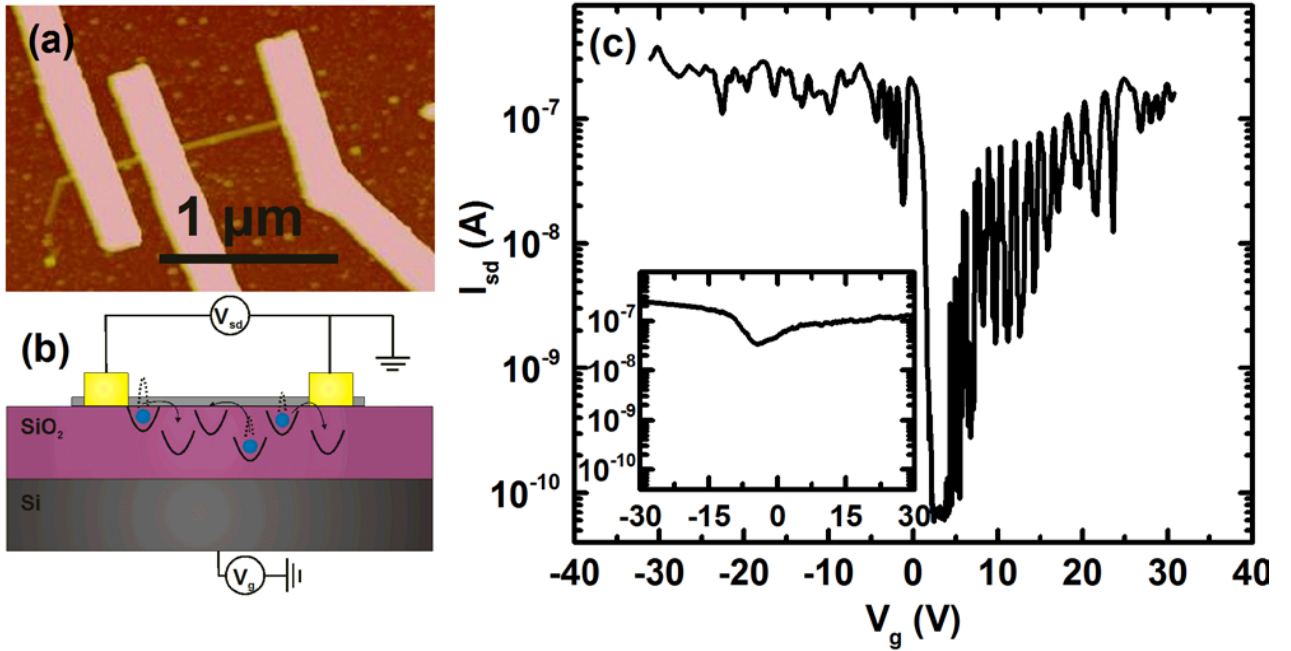

**Figure 1.** (a) AFM image of the typical sample. (b) Schematic drawing of the device. Charge traps fluctuate (trapping and detrapping) and can induce a potential barrier affecting the nanotube channel. Also measurement setup is drawn here. (c) Typical gate curve of IS-MWCNT device in 4.2 K revealing the gap opening compared to the room temperature curve (inset).

In figure 1 we show the transport characteristics ($I_{sd} - V_g$) of a typical Pd contacted semiconducting MWCNT FET at 4.2 K. All devices studied here show a negligible gap in the $I_{sd} - V_g$ characteristics at room temperature but a clear gap emerges on cooling down. The examined devices are ambipolar in nature with an ON/OFF-ratio around four orders of magnitude. Considerable oscillations in the source-drain current $I_{sd}$ with gate voltage reminiscent of Fabry-Perot type interferences in the p-side with oscillations of around $(2-3) \cdot e^2\hbar$ [9] and Coulomb blockade oscillations in the OFF-state for the n-side can be observed. Fabry-Perot interferences are a signature of good quality tubes (ballistic channel) due to the quantum interference at the contacts. In contrast, Coulomb blockade effects in the n-side are a result of weaker contacts perhaps due to larger Schottky barrier between metal and conduction band of the tube [9, 28]. Different transparencies for the n- and p-type regions thus exist in the MWCNTs.

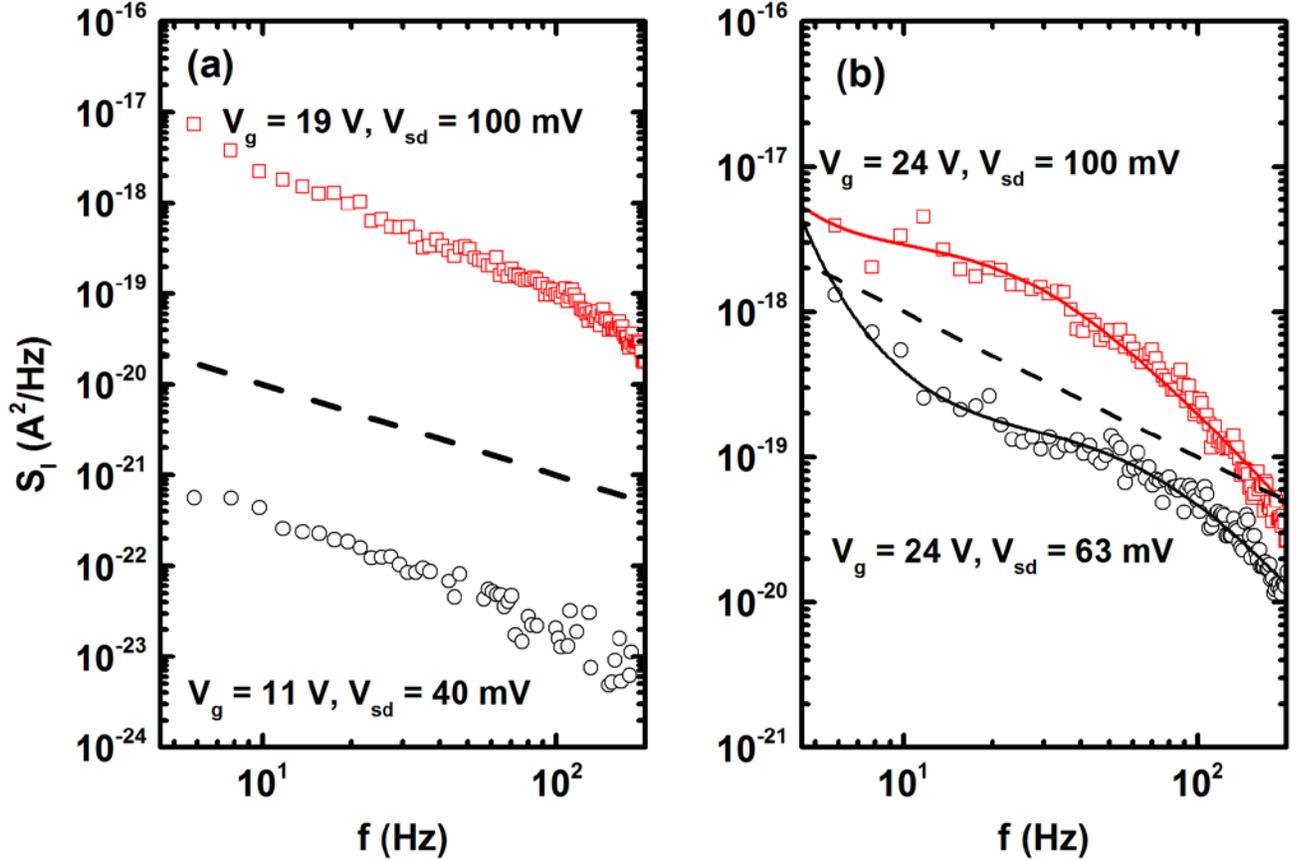

**Figure 2.** Current power spectrum $S_I$ as a function of frequency for tube A at various source-drain, $Vs_d$ voltages and gate voltages, $V_g$. (a) Noise spectra following nicely 1/f behavior (indicated by the dashed line). (b) Noise spectra revealing lorentzian line shapes. The continuous curves are lorentzian fits to the data.

Typical results of the noise measurement (tube A) are shown in figure 2. The current power spectral density (PSD) of noise $S_I(f)$ at different values of $V_g$ and $V_{sd}$ are presented. The spectral density was found to be a mix of 1/f and two level fluctuations (RTN) for different values of source-drain bias $V_{sd}$. For some values of $V_g$ and $V_{sd}$ the spectra can also be Lorentzian (2 (b)) which will be discussed later. The time series of current was recorded in a 250 Hz bandwidth and the PSD of noise was found to have a quadratic dependence on the applied bias.

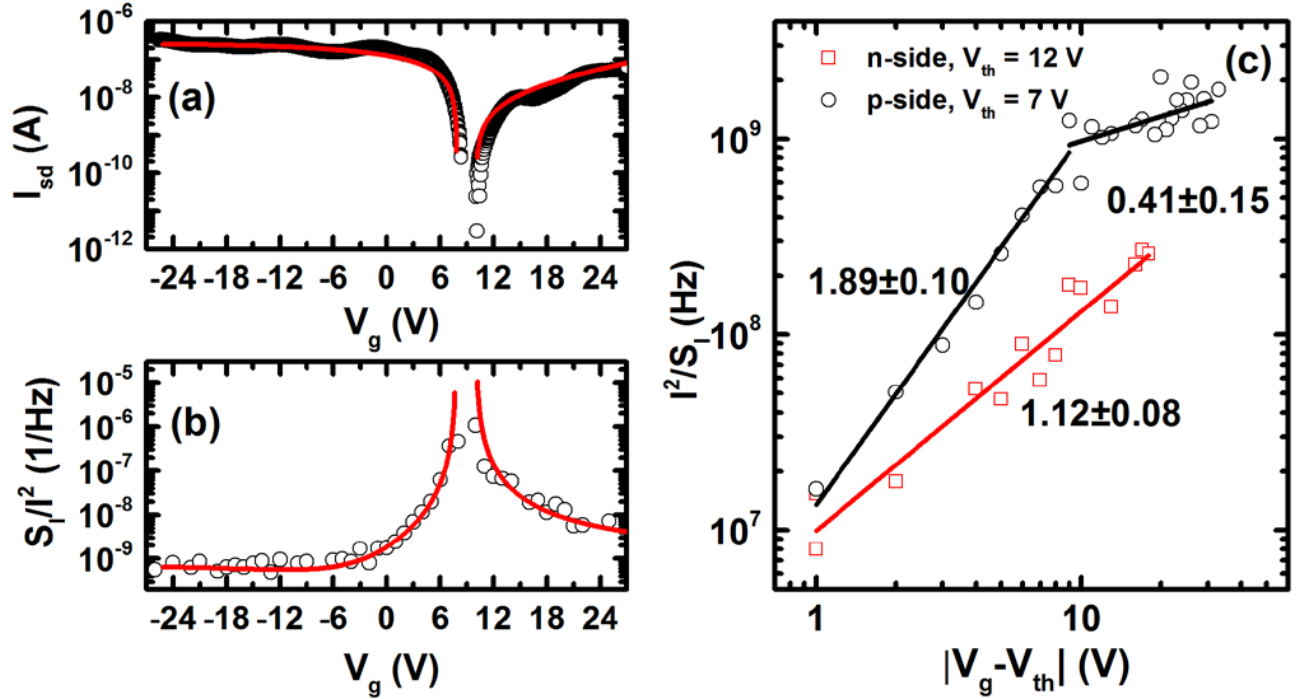

**Figure 3.** (a) Smoothed source-drain current as a function of gate voltage for tube A (dots) and exponential fit to the data (red curve). Gate curve is measured with 1 mV source-drain voltage. (b) Normalized noise amplitude vs. gate voltage for the same device. Circles are the experimental data and solid lines are the fits made by using the charge noise model. (c) Inverse of normalized noise vs. $|V_g - V_{th}|$ in log-log scale for p-side (black dots, $Vt_h$ = 7 V) and n-side (red squares, $V_{th}$ = 12 V). Solid lines are linear fits to the data revealing two different exponents in the p-side (1.89 & 0.41) and exponent value of 1.12 in the n-side.

Figure 3 (a) shows the transfer characteristics $I_{sd} - V_g$ for tube A of channel length 130 nm and diameter 7.1 nm. The device has an ON-state resistance value of $R_{ON}$ = 6.2 kΩ being close to ballistic limit. The functional dependence of $I_{sd}(V_g)$ is exponential in the subthreshold regime for all the studied devices indicated by the solid red line for tube A. The device has a subthreshold slope of 0.8 V/decade in the p-region and 3.1 V/decade in the n-region. The current noise power spectrum ($S_I(f)$) of the device was also recorded as a function of the gate voltage. Figure 3 (b) shows the normalized noise amplitude ($S_I/I^2$) as a function of gate voltage $V_g$ for the same device. The normalized noise has been found for all the devices by averaging over 20 − 40 Hz frequency octave. It can be clearly seen from the figure that the normalized noise is minimum in the ON-state and reaches a maximum while going up to the subthreshold region (OFF-state). This behavior is quite similar to those reported by Lin et. al. [1] for SWCNTs. However, unlike SWCNTs the normalized noise is one to two orders of magnitude lower than SWCNTs [1-4] and one order lower than previously studied quasi-ballistic MWCNTs [23]. The lower noise in MWCNT-FETs might be owing to larger diameter of the tubes as the sensitivity of the current through the CNT channel on individual traps is reduced compared to SWCNTs.

We now analyze the gate dependence of noise within the purview of McWhorter model [4, 13] which is essentially an interface dominated model. Here, the noise magnitude is used to discriminate between carrier-number fluctuations $\delta N$ and mobility fluctuations $\delta \mu$ by studying its gate voltage dependence. The inverse normalized noise ($I^2/S_I$) is studied as a function of $|V_g - V_{th}|^2$ where $V_{th}$ is the threshold voltage. For number fluctuations $\delta N$, $I^2/S_I = D \ |V_g - V_{th}|^2$ and $I^2/S_I = D \ |V_g - V_{th}|$ for mobility fluctuations. $D$ is a constant given by $C_g L_C/\alpha_h e$, where $C_g$, $L_C$, $\alpha_h$ and $e$ are the gate capacitance, channel length, Hooge's constant and electronic charge, respectively. A log-log plot of ($I^2/S_I$) with $|V_g - V_{th}|$ can provide an approximation of the exponent of $|V_g - V_{th}|$ as well as determine the applicability of this model. From figure 3 (c) it is quite clear from the p-side noise behavior that there are two different exponents of $|V_g - V_{th}|$ i.e., 1.89 ±0.10 from 1 to 10 V and 0.41 ±0.15 above that which is in clear disagreement with McWhorter model. As the transport in the tubes is ballistic, number fluctuations can also be ruled out as the origin of noise even though the exponent of the low voltage region is close to 2. Disagreement with McWhorter model is also evident in the high voltage regime as the exponent of 0.41±0.15 deviates significantly from the expected values. For the n-side the noise seems to go linearly with $|V_g - V_{th}|$ with an exponent of 1.12 ±0.08 pointing towards mobility fluctuations as possible cause for producing the fluctuations. However, computing the value of $D$, assuming [2] $\alpha_h \sim 10^{-2}$ and other parameters as in Table I, gives a value of $\sim 10^{-6}$ which is many orders of magnitude different from the experimentally obtained values of $1.5 \times 10^7$ for n-side and $5 \times 10^7$ for the p-side. The McWhorter's model which has its origin in Hooge's law is thus unable to explain the noise data in these devices which is not surprising in this case as the transport in the channel is ballistic. The conclusions are further supported by the independence of the normalized noise with the channel length of the devices given in Table I later.

In light of the failure of McWhorter's model to explain the noise results we analyze the gate dependent noise behavior with the charge noise model as proposed by Tersoff [12] more suitable for ballistic devices. In nanoscale FETs, the fluctuations from trap states in the oxide is responsible for modulating the quantum transmission of the channel for ballistic transport. The normalized noise for charge noise model is generally written as

$$A_I = \gamma^2 S_g^2 \left(\frac{d \ln I_{sd}}{dV_g}\right)^2 + \alpha_c I_{sd}^2, \quad (1)$$

where the first term is the charge noise term in the subthreshold region. In the crucial subthreshold region low frequency noise originates from the contact in the Schottky barrier devices or from the channel itself in purely channel limited ones. The second term is the noise in the ON-state represented as a classical series resistor independent from the gate effects near the contacts having a noise of $\alpha_c I_{sd}^2$, where $\alpha_c = \delta R_c^2/V_{sd}^2$, $\delta R_c^2$ being the resistance noise power of the series resistor. Parameter $S_g$ describes the device geometry and $\gamma$ the quality of the gate oxide. In the case of Schottky barriers (SB) $\gamma$ describes the number of charge traps and their proximity to the contact. When there is no SBs present e.g. a channel limited CNT-FET, $\gamma$ represents the fluctuation of maximum potential along the channel. We first deduce the charge noise term $\gamma^2 S_g^2 (d \ln I_{sd}/dV_g)^2$ of equation 1 by fitting the experimental $I_{sd}(V_g)$ data with an exponential function and differentiating that analytically. The $\gamma S_g$ value for the tube A (in fig. 3 (b)) is 0.5 meV. Note that the fitting has to be done separately for $p-$ and $n-$ side in order to get good exponential dependence. This means there is a small gap in the data at the truly OFF-state. In order to get a good correspondence to the data in the ON-regime as well a second term $\alpha_c I_{sd}^2$ has to be added to the charge noise expression. The second term represents the noisy resistance connected in series to the channel. Using $\alpha_c$ as another fitting parameter we get excellent fitting throughout the whole gate range. As seen from figure 3 (b), the charge noise model accurately explains the noise behavior in the entire region including the *p*- region where the transport is close to ballistic and the higher Schottky barrier *n-side*.

**Table 1.** Diameter *d*, channel length $L_C$, ON-state resistance $R_{ON}$, fitting parameter $\gamma S_g$ (p-side), fitting parameter $\alpha_c$ (p-side) and gate capacitance $C_g/L$ per unit length for four measured devices and one from ref. 1.

|        | *d* (nm) | $L_C$ (nm) | $R_{ON}$ (kΩ) | $\gamma S_g$ (meV) | $\alpha_c$ (1/Hz) | $C_g/L$ (pF/m) |
|--------|----------|------------|---------------|--------------------|-------------------|----------------|
| Tube A | 7.1      | 130        | 6.2           | 0.5                | 1x10⁴             | 42             |
| Tube B | 6.0      | 420        | 18.4          | 0.6                | 5x10⁵             | 41             |
| Tube C | 3.0      | 390        | 15.0          | 0.3                | 4x10⁵             | 36             |
| Tube D | 7.1      | 570        | 10.0          | 0.3                | 5x10⁴             | 42             |
| Ref. 1 | 1.8      | 600        | ~10           | 7.0                | -                 | 70             |

The normalized noise as a function of gate voltage for all the four devices were fitted to the charge noise model and the values of $\gamma S_g$ extracted from the fits and other parameters are given in Table I. The obtained values of the noise parameter $\gamma S_g$ extracted for the IS-MWCNT devices is one order of magnitude smaller than the values obtained for SWCNTs from Lin's data (see ref. 12). The noise parameter $\gamma S_g$ is related [14] to the total gate capacitance as $\gamma S_g \propto (1/C_g)^2 S_q$, where $S_q$ is the charge distribution fluctuation. The gate capacitance for the tubes can be estimated by considering the capacitance between a plate and a wire, which have the capacitance per unit length as $C_g/L_c = 2\pi E_0 E_r / \ln(4h/d)$, where $E_r$ is the dielectric constant, $h$ the thickness of the gate oxide and $d$ the diameter of the tube. The capacitance of all devices used in this work and from Lin's paper is shown in Table I. The capacitance per unit length of the SWCNT device is found to be more than twice of MCWNT values however they do not vary much among themselves. As the length dependence of total capacitance is $C_g \propto L_c$ and the charge distribution $S_q \propto L_c$ due to uniform distribution of fluctuators along the channel, the noise parameters should have length dependence as $\gamma S_g \propto 1/L_c$ where $L_c$ is the channel length. However, we do not obtain any channel length dependence of noise as obtained in liquid gated SWCNT devices [14] and graphene transistors [29]. In these works the length of the channel was varied sufficiently going up to the diffusive regime. However, as the scope of the charge noise model is limited to ballistic transport regime only we carried out the experiments only for tubes with short channel lengths.

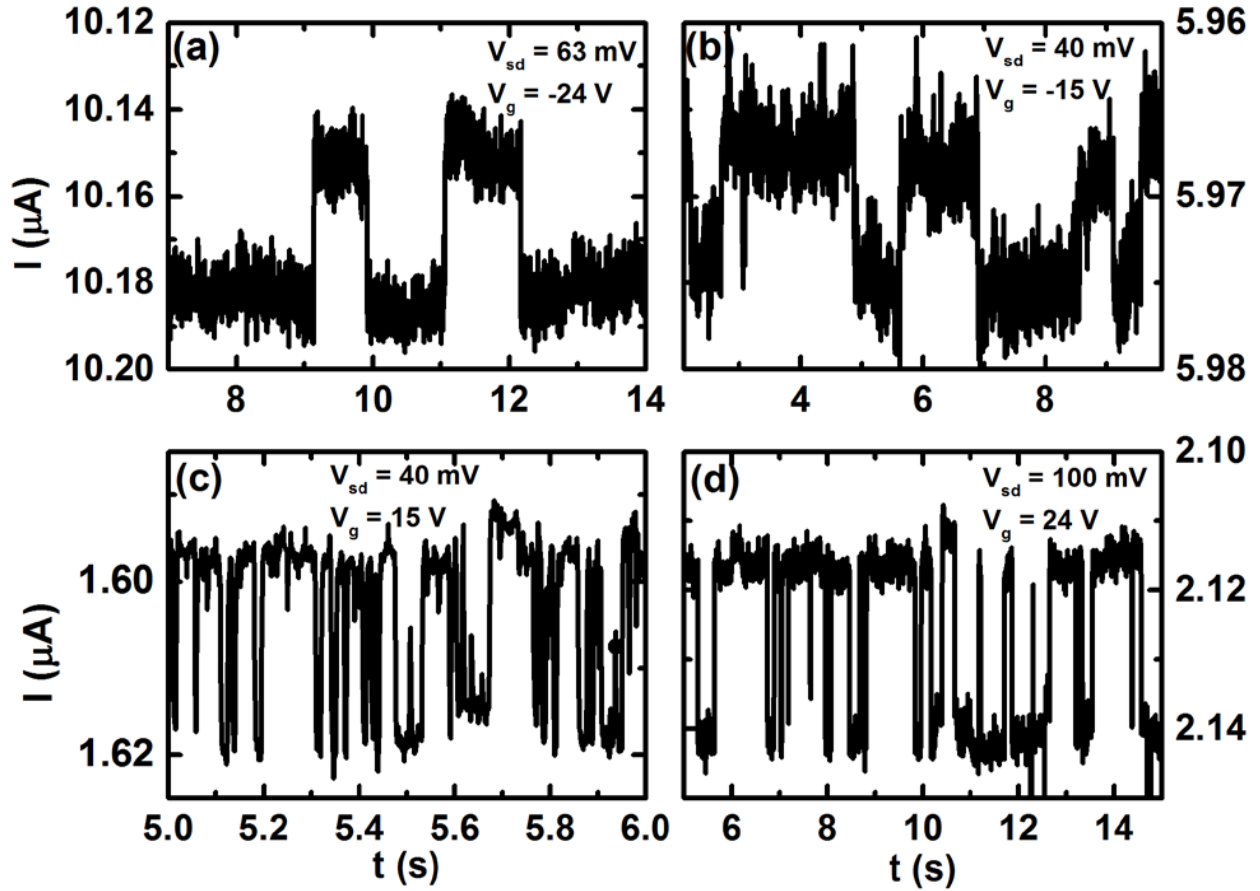

Figure 4. RTN signals of nanotube A in different gate and source-drain voltages. The data shows random and independent configuration of RTN in respect to $V_g$ and $V_{sd}$. RTN in (a) on-state (p-side) ($Vg = -24$ V and $Vsd = 63$ mV), (b) closer to subthreshold region ($Vg =$

−15 V and $V_{sd}$ = 40 mV), (c) subthreshold region in n-side ($V_g$ = 15 V and $V_{sd}$ = 40 mV) (d) ON-state (n-side) ($V_g$ = 24 V and $V_{sd}$ = 100 mV).

The validity of the charge noise model can also be checked by studying the random telegraph signal noise (RTN) obtained for the CNT-FETs. RTN is common in SWCNT devices and have been studied both experimentally [30] and theoretically [31] whose origin is attributed to the trapping and detrapping of charges in the oxide. However in some earlier studies the origin of RTN for large diameter MWCNTs has also been ascribed to fluctuations from the contacts [19]. Usually RTN is characterized by a lorentzian [32, 33] power spectra ($S_I(\omega) = 4I^2\tau_{eff}^2/(1 + \omega^2\tau^2)$) and a characteristic time given by $1/\tau_1 + 1/\tau_2$, where $\tau_1$ and $\tau_2$ are the capture and emission times respectively. The characteristic time reduces exponentially with applied bias [34]. At high bias the spectra thus shows an evolution from the lorentzian to $1/f$ type. However, for all the MWCNT devices studied in this work, the experimental results point to no particular $V_{sd}$ or $V_g$ dependence of RTN. In figure 4, we plot a few typical RTN at four random $V_{sd}$ and $V_g$ values for the tube. It can be clearly seen from the figure that the current variation varies between 0.5−5 % which is similar to RTN values obtained in conventional silicon MOSFETs but lower than SWCNTs where values as high as 50 % have been reported [31] which can persist even in the ON-state. The presence of RTN in the ON-state is also present for the IS-MWCNTs devices studied here as illustrated in figure 4 (a) and (d). The noise behavior in IS-MWCNTs is qualitatively similar to SWCNTs though with lower noise magnitude which is understandable as SWCNTs have smaller diameters and are thus more sensitive to RTN effects than larger diameter MWCNT devices. The modulation of transmission probability in ballistic IS-MWCNTs is thus quite similar to that in ballistic SWCNTs implying strong oxide trap modulations pointing towards the applicability of charge noise model.

## 4. CONCLUSION

In conclusion, we have presented the low frequency noise properties for intermediate sized ballistic semiconducting MWCNT-FETs. The gate dependence of low frequency noise in these devices can be better described using a charge noise model instead of the more traditional McWhorter's model. Our experiments point towards the fact that the crucial factor for lowering/optimizing noise properties of ballistic FETs is by controlling the diameter of tubes and the quality of gate oxide. It is imperative that a fully suspended structure with vacuum as dielectric would result in better low noise transistors. These findings have important implications to the noise literature of carbon nanotubes bridging the gap between SWCNTs and large MWCNTs and to the noise properties of ballistic nanoscale transistors in general.

## 5. ACKNOWLEDGEMENTS

The authors would like to acknowledge Jenny and Antti Wihuri Foundation, Finnish Cultural Foundation (O.H.) and CIMO (D.T.) for supporting this work.

## REFERENCES

[1] Lin Y-M, Appenzeller J, Knoch J, Chen Z, Avouris P. Low-Frequency Current Fluctuations in Individual Semiconducting Single-Wall Carbon Nanotubes. Nano Letters 2006; 6(5):930-36.

[2] Tobias D, Ishigami M, Tselev A, Barbara P, Williams ED, Lobb CJ, Fuhrer MS. Origins of 1/f noise in individual semiconducting carbon nanotube field-effect transistors. Phys. Rev. B 2008; 77(3):033407.

[3] Liu F, Wang KL, Zhang D, and Zhou C. Noise in carbon nanotube field effect transistor. Appl. Phys. Lett. 2006, 89(6):063116.

[4] Ishigami M, Chen JH, Williams ED, Tobias D, Chen YF and Fuhrer MS. Hooge's constant for carbon nanotube field effect transistors. App. Phys. Lett. 2006; 88(20):203116.

[5] Biercuk MJ, Ilani S, Marcus CM, McEuen PL. Electrical Transport in Single-Wall Carbon Nanotubes. In: Jorio A, Dresselhaus MS, Dresselhaus G, editors. Topics in Applied Physics, vol 111, Berlin; Springer; 2008, p. 455–493.

[6] Ahlskog M, Herranen O, Johansson A, Leppaniemi J, Mtsuko D. Electronic transport in intermediate sized carbon nanotubes. Phys. Rev. B 2009; 79(15):155408.

[7] Cumings J, Zettl A. Localization and Nonlinear Resistance in Telescopically Extended Nanotubes. Phys. Rev. Lett. 2004; 93(8):086801.

[8] Javey A, Guo J, Lundstrom M, Dai H. Ballistic carbon nanotube field-effect transistors. Nature 2003; 424(6949):654-57.

[9] Liang W, Bockrath M, Bozovic D, Hafner J, Tinkham M, Park H. Fabry - Perot interference in a nanotube electron waveguide. Nature 2001; 411(6838):665-69.

[10] Kong J, Yenilmez E, Tombler TW, Kim W, Liu L, Jayanthi CS, Wu SY, Laughlin RB, Dai H. Quantum Interference and Ballistic Transmission in Nanotube Electron Waveguides. Phys. Rev. Lett. 2001; 87(10):106801.

[11] Mann D, Javey A, Kong J, Wang Q, Dai H. Ballistic Transport in Metallic Nanotubes with Reliable Pd Ohmic Contacts. Nano Letters 2003; 3(11):1541-44.

[12] Tersoff J. Low-Frequency Noise in Nanoscale Ballistic Transistors. Nano Letters 2007; 7(1):194-98.

[13] McWhorter AL. 1/f Noise and Germanium Surface Properties. In: Kingston RA, editor. Semiconductor Surface Physics, Philadelphia: University of Pennsylvania Press; 1957, p. 207–228.

[14] Männik J, Heller I, Janssens AM, Lemay SG, Dekker C. Charge Noise in Liquid-Gated Single-Wall Carbon Nanotube Transistors. Nano Letters 2008; 8(2):685-88.

[15] Balandin AA. Low-frequency 1/f noise in graphene devices. Nature Nanotechnology 2013; 8(8):549-55.

[16] Lassagne B, Raquet B, Broto J-M, Cleuziou J-P, Ondarcuhu Th, Monthioux M, Magrez A. Electronic fluctuations in multi-walled carbon nanotubes. New J. Phys. 2006; 8(3):31

[17] Roschier L, Tarkiainen R, Ahlskog M, Paalanen M, Hakonen P. Multiwalled carbon nanotubes as ultrasensitive electrometers. Appl. Phys. Lett. 2001; 78(21):3295-97

[18] Ouacha H, Willander M, Yu HY, Park YW, Kabir MS, Persson SHM, Kish LB, Ouacha A. Noise properties of an individual and two crossing multiwalled carbon nanotubes. Appl. Phys. Lett. 2002; 80(6):1055-57

[19] Tarkiainen R, Roschier L, Ahlskog M, Paalanen M, Hakonen P. Low-frequency current noise and resistance fluctuations in multiwalled carbon nanotubes. Physica E 2005; 28(1):57-65.

[20] Bachtold A, Strunk C, Salvetat J-P, Bonard J-M, Forro L, Nussbaumer T, and Schönenberger C. Aharonov–Bohm oscillations in carbon nanotubes. Nature 1999; 397(6721):673-75.

[21] Wei BQ, Vajtai R, Ajayan PM. Reliability and current carrying capacity of carbon nanotubes. Appl. Phys. Lett. 2001; 79(8):1172-74.

[22] Li H, Lu WG, Li JJ, Bai XD, Gu CZ. Multichannel Ballistic Transport in Multiwall Carbo Nanotubes. Phys. Rev. Lett. 2005; 95(8):086601.

[23] Bourlon B, Miko C, Forró L, Glattli DC, Bachtold A. Determination of the Intershell Conductance in Multiwalled Carbon Nanotubes. Phys. Rev. Lett. 2004; 93(17):176806

[24] Zhou C, Kong J, Dai H. Intrinsic Electrical Properties of Individual Single-Walled Carbon Nanotubes with Small Band Gaps. Phys. Rev. Lett. 2000; 84(24):5604-07.

[25] Koshio A, Yudasaka M, Iijima S. Metal-free production of high-quality multi-wall carbon nanotubes, in which the innermost nanotubes have a diameter of 0.4 nm. Chem. Phys. Lett. 2002; 356(5-6):595-600.

[26] Talukdar D, Yotprayoonsak P, Herranen O, Ahlskog M. Linear current fluctuations in the power-law region of metallic carbon nanotubes. Phys. Rev. B 2013; 88(12):125407.

[27] Talukdar D, Chakraborty RK, Bose S, Bardhan KK. Low noise constant current source for bias dependent noise measurements. Rev. of Sci. Instr. 2011; 82(1):013906

[28] Grove-Rasmussen K, Jorgensen HI, Lindelof PE. Fabry–Perot interference, Kondo effect and Coulomb blockade in carbon nanotubes. Physica E 2007; 40(1):92-8.

[29] Heller I, Chatoor S, Männik J, Zevenbergen MAG, Oostinga JB, Morpurgo AF, Dekker C, Lemay

SG. Charge Noise in Graphene Transistors. Nano Letters 2010; 10(5):1563-67.

[30] Liu F, Bao M, Kim H, Wang KL, Li C, Liu X, Zhou C. Giant random telegraph signals in the carbon nanotubes as a single defect probe. Appl. Phys. Lett. 2005; 86(16):163102.

[31] Wang N-P, Heinze S, Tersoff J. Random-Telegraph-Signal Noise and Device Variability in Ballistic Nanotube Transistors. Nano Letters 2007; 7(4):910-13.

[32] Malchup S. Noise in Semiconductors: Spectrum of a Two-Parameter Random Signal. J. Appl. Phys. 1954; 25(3):341-43.

[33] Hung KK, Ko PK, Hu C, Cheng YC. Random Telegraph Noise of Deep-Submicrometer MOSFET's. IEEE electron device letters 1990; 11(2):90-2.

[34] Uren MJ, Day DJ, Kirton M. 1/f and random telegraph noise in silicon metal-oxide-semiconductor field-effect transistors. Appl. Phys. Lett. 1985; 47(11):1195-97.